\documentclass[%
 jmp,
amsmath,amssymb,
preprint,%
]{revtex4-1}

\usepackage{graphicx}
\usepackage{dcolumn}
\usepackage{bm}
\usepackage{amsmath,amsfonts,amssymb,graphicx}
\usepackage{mathrsfs,calrsfs}
\usepackage{etoolbox}
\usepackage{color}
\usepackage{cancel}
\usepackage{soul}
\DeclareMathAlphabet{\pazocal}{OMS}{zplm}{m}{n}
\newcommand{\Pb}{\pazocal{P}}
\newcommand{\Tb}{\pazocal{T}}
\newcommand{\PTb}{\pazocal{PT}}

\usepackage[normalem]{ulem}
\newcommand{\bq}{\begin{equation}}
	\newcommand{\ba}{\begin{eqnarray}}
		\newcommand{\eq}{\end{equation}}
	\newcommand{\ee}{\end{equation}}
\newcommand{\ea}{\end{eqnarray}}

\makeatletter
\def\@email#1#2{%
\endgroup
\patchcmd{\titleblock@produce}
{\frontmatter@RRAPformat}
{\frontmatter@RRAPformat{\produce@RRAP{*#1\href{mailto:#2}{#2}}}\frontmatter@RRAPformat}
{}{}
}%
\makeatother

\begin{document}
	
\preprint{AIP/123-QED}
	
	\title[]
	{Generalized  $\PTb$-symmetric nonlinear Dirac equation: exact solitary waves solutions, stability and conservation laws}
\author{Fernando Carreño-Navas}
\email{fcarreno@us.es} 
\affiliation{IMUS, Universidad de Sevilla, 41012, Spain}	
\affiliation{Departamento de Análisis Matemático, Universidad de Sevilla, c/Tarfia s/n, Sevilla, 41012, Spain}
\author{Siannah Peñaranda}
\email{siannah@unizar.es}
\affiliation{Departamento de Física Teórica and Centro de
  Astropartículas y Física de Altas Energías (CAPA), Universidad de
  Zaragoza, Pedro Cerbuna 12, Zaragoza 50009, Spain}
\author{Renato Alvarez-Nodarse}
\email{ran@us.es} 
\affiliation{Departamento de Análisis Matemático, Universidad de Sevilla, c/Tarfia s/n, Sevilla, 41012, Spain}
\affiliation{IMUS, Universidad de Sevilla, 41012, Spain}
\author{Niurka R.\ Quintero}
\affiliation{Departamento de F\'\i sica Aplicada I, ETSII,  Universidad de Sevilla,
Avda Reina Mercedes s/n, Sevilla, 41012, Spain}	
\email{niurka@us.es}
\affiliation{IMUS, Universidad de Sevilla, 41012, Spain}
\date{\today}

\begin{abstract}
We derive an exact solitary wave solution for the $\PTb$-symmetric nonlinear Dirac equation with a scalar-scalar interaction. We consider a power-law nonlinearity of the form $|\bar{\Psi}\,\Psi|^{k}\,\Psi$ for positive values of $k$. 	
The system's energy is conserved despite the presence of a gain-loss term, which is quantified by the parameter $\Lambda$. We show  that the $\PTb$-transition point is defined by the solution's existence condition and is independent of the nonlinearity exponent $k$. 
Furthermore, momentum is conserved, although neither the canonical momentum nor the charge is a conserved quantity. A notable result is that the stationary solution, obtained from the continuity equations, exhibits nonzero momentum in its rest frame. We also derive a moving soliton solution, where the gain-loss parameter allows the soliton's velocity to be precisely chosen so that the moving soliton possesses zero momentum. Finally, we establish that the presence of a gain-loss mechanism and higher-order nonlinearity restrict the stability domain of the solutions.
\end{abstract}

\maketitle

\section{Introduction\label{intro}} 

The study of solitary wave solutions in the (1+1)-dimensional nonlinear
Dirac (NLD) equation has a long and significant history. The pioneering
investigations of this topic were established through seminal works on
the Thirring model~\cite{thirring:1958} and the Gross-Neveu (GN)
equation~\cite{gross:1974}. These models represent the classical
equations of motion for field theories of massive fermions with Fermi
interactions in one space and one time dimension. Specifically, in 1970
Soler~\cite{soler:1970} proposed  the solitary waves of an NLD field
with self-interacting spinors (GN) as a model of elementary
fermions. Subsequent research has focused on establishing a number of
fundamental results concerning the exact solitary wave solutions and the
methods for their
derivation~\cite{lee:1975,mathieu:1985,merle:1988}. For example, the
Thirring equation was solved using the inverse scattering transform,
which established its integrability~\cite{mikhailov:1976,kaup:1977},
while other methodological approaches have included the use of
conservation laws and variational approach to obtain analytical
solutions~\cite{chang:1975,mathieu:1985,nogami:1992,cooper:2010}.  
These traditional methods can also be applied to find solutions 
for  $\PTb$-symmetric ($\Pb$ represents parity reflection and $\Tb$
represents time reversal) NLD equations~\cite{alexeeva:2019}, even in
the presence of dissipation.  

Even if the NLD equations were originally proposed in the context
of particle physics, the discovery that non-Hermitian but
$\PTb$-symmetric Hamiltonians can possess real
eigenvalues~\cite{bender:1998,bender:2002,bender:2005,bender:2007}
has stimulated the study of $\PTb$-symmetric systems across numerous
fields, from linear systems to nonlinear
physics~\cite{musslimani:2008,guo:2009,ruter:2010,barashenkov:2013,tsironis:2014,konotop:2016,cheong:2024}. Non-Hermitian
quantum mechanics can be emulated by optical media with the
refractive index or permittivity~\cite{konotop:2016}. While
$\PTb$-symmetric models are commonly formulated by introducing a
complex
potential~\cite{kevrekidis:2015,yan:2017,nath:2017,mertens:2024},
Barashenkov et al. specifically investigated a class of these
$\PTb$-symmetric models that admit exact soliton
solutions~\cite{barashenkov:2015}.

The concept of $\PTb$  symmetry found a powerful application in overcoming a common issue in
dissipative systems: the eventual disappearance of damped solitons. Specifically, in nonlinear
optics, the use of $\PTb$-symmetric  optical lattices, which employ a balanced distribution of gain
and loss elements, was shown to support and stabilize soliton solutions~\cite{longhi:2009,alexeeva:2012,tsoy:2014}. This same phenomenology, using a gain-loss parameter, has also been successfully
employed to obtain stable solitons in $\PTb$-symmetric variants of the
Thirring, Gross-Neveu and Alexeeva-Barashenkov-Saxena
equations~\cite{alexeeva:2019}. Unlike those in generic dissipative systems, these particular
solutions  exist, often possess conserved quantities, and preserve Lorentz invariance. This
phenomenology contrasts with the parametrically driven and damped NLD equation, where the
stationary solitary wave emerges from a balance between the damping and the parametric force,
with both acting identically in each spinor~\cite{quintero:2019b,sanchez:2025}.

The interplay between the nonlinearity $(\bar{\Psi}\,\Psi)^{k}\,\Psi$
and the $\PTb$-symmetry induced by including a gain-loss term
proportional to the parameter $\Lambda \ne 0$ that involves the Dirac
matrix $\gamma^5$ was proposed in~\cite{cuevas:2016a}. The
$\PTb$-symmetric GN model was considered, by giving a detailed numerical
simulations for $k=1$. The authors of Ref. \cite{cuevas:2016a}
demonstrated that, although energy was conserved, charge oscillated
due to initial conditions and a source term in the continuity equation.
Moreover, the $\PTb$ transition was numerically identified at
$\Lambda=\Lambda_{\PTb}=\sqrt{m^2-\omega^2}$ being $\omega$ the
frequency of stationary solution, when the nonlinear solution collided
with linear waves. Exact analytical solutions were identified only for
the massless case $m=0$. Notably, a recent investigation of the $k=1$
case in~\cite{alexeeva:2019} identified a solitary wave and established
the conservation of energy and momentum. Furthermore, Khare et al. \cite{khare:2026} recently discovered a two-parameter family of generalized nonlinear Dirac solitary waves in $1+1$ dimensions, characterized by scalar-scalar and vector-vector interactions (generalized Alexeeva-Barashenkov-Saxena model).

Crucially, we now introduce a modification in the generalized nonlinear
term (proportional to $|\bar{\Psi}\,\Psi|^{k}\,\Psi$) designed to ensure
a strictly real energy functional, thereby avoiding the complex energy
inconsistencies inherent in earlier models. In the current work, we show
that solitary wave solutions exist for all $k>0$, instead of only $k =
1$. We have related the critical parameter $\Lambda$ with the condition
for the existence of stationary solution. Furthermore, for all $k>0$,
not only the energy and the momentum are conserved, but the charge and
the canonical momentum associated with the stationary solution are
conserved as well. Additionally, by employing the conservation laws of
energy and momentum (see Ref.~\cite{alexeeva:2019}), along with a set
of identities involving (i) the energy density and momentum flux, and
(ii) the momentum density and energy flux, we derive the exact solitary
wave solution of the system of two coupled nonlinear partial
differential equations obtained from the $\PTb$-symmetry GN model in
covariant form for all $k>0$. We also obtain the exact expressions of the energy,
momentum, and the charge of the stationary solution and moving solitary
wave, showing their dependence on $\Lambda$ and on the frequency of
stationary solution. Finally, we demonstrate that for $k>2$, there
exists a specific region in the $(\Lambda,\omega)$ parameter space
where the soliton becomes unstable. Conversely, numerical evidence
points to stable solitons for all $k \le 2$, even though analytically, 
the stationary solution only meets the necessary stability condition
established by the Vakhitov-Kolokolov criterion.

The outline of the paper is as follows. In Section~\ref{sec2}, we introduce   
the generalized $\PTb$-symmetric Gross-Neveu model and delineate its 
main symmetry properties. Subsequently, in Section~\ref{sec3} we derive the exact
stationary solution and we obtain moving solitons by applying the
Lorentz boost. Furthermore, we calculate the conserved quantities
associated with these solutions. In Section~\ref{sec4}, we perform a
stability analysis of stationary solution. Main results and concluding
remarks are summarized in Section~\ref{sec5}.

\section{$\PTb$-symmetric Gross-Neveu model}
\label{sec2}

The $\PTb$-symmetric Gross-Neveu model in covariant form is given by 
\begin{align}
\label{eq0}
i\,\gamma^{\mu}\,\partial_{\mu}
  \Psi-m\,\Psi+g\,|\bar{\Psi}\,\Psi|^{k}\,\Psi&
  =\gamma^5\,\Lambda \,\Psi,
\end{align}
where the spinor field is defined as $\Psi(x,t)=(U(x,t),
V(x,t))^{\scriptscriptstyle T}$ with $U(x,t)$ and $V(x,t)$ denoting the
two complex spinor components,   
 $\bar{\Psi}=(\Psi^\star)^{\scriptscriptstyle T}\,\gamma^0$ is the
 adjoint spinor, $\gamma^{\mu}$ are the Dirac matrices, $m$ is the mass parameter 
 and $g$ represents the coupling constant that characterizes the scalar
 fermion self-interactions. The parameters $m$, $g$, and $k$ are
 positive coefficients, and the matrix
 $\gamma^5$ is defined as  $\gamma^{5}=\gamma^0\,\gamma^1$, where 
\begin{align} 
\label{m1}
\gamma^{0} &= 
\left( \begin{array}{cc}
	1 & 0  \\
	0 & -1  \end{array} \right), \qquad \gamma^{1} = 
\left( \begin{array}{cc}
	0 & i  \\
	i & 0  \end{array} \right)\,.
\end{align}
This model is modified from the one proposed in Ref.~\cite{cuevas:2016a}
in order to study the interplay between the nonlinearity and the  $\PTb$-symmetry, induced by
including a gain-loss term proportional to the parameter $\Lambda \ne 0$. In this case, the
term $|\bar{\Psi}\,\Psi|^{k} \Psi$ has been introduced to model how nonlinearity affects the
behavior of the solutions and to describe the system as a real Lagrangian with a dissipation
function (external force), $\mathcal{F}$, as
\begin{align}
	\label{Lagrangian}
	\mathcal{L} &= \frac{i}{2} [\bar{\Psi}
                      \,\gamma^{\mu}\,\partial_{\mu} \Psi
                      -\partial_{\mu}\bar{\Psi}\,\gamma^{\mu}\,\Psi]
                      -m\,\bar{\Psi}\,\Psi+\frac{g}{k+1}\,|\bar{\Psi}\,\Psi|^{k}\,(\bar{\Psi}\Psi)
                      ,\\ 
	\label{External_Force}
	\mathcal{F} &=\Lambda \,(\partial_t \bar{\Psi}
                      \,\gamma^5\,\Psi-\bar{\Psi} \,\gamma^5\,\partial_t
                      \Psi)\,. 
\end{align}
The first two terms in $\mathcal{L}$ correspond to the standard Dirac
kinetic term and the third one to the Dirac mass term. The last term in
this Lagrangian corresponds to the nonlinear self-interaction
of the fermion field  and generalizes models such as the Gross-Neveu model for $k=1$. Here, $g$
encodes the strength of the interaction, and $k$ encodes the degree of
nonlinearity. If $k=1$ one has a quartic fermion interactions. 

Eq.~\eqref{eq0} is equivalent to the following system of two coupled partial differential equations:
\begin{align}
	\label{eq1}
	i U_t &= \, V_x-g\,||U|^2-|V|^2|^{k} U+m\,U+i\,\Lambda\,V, \\
	\label{eq2}
	i V_t &= -U_x+g\,||U|^2-|V|^2|^{k} V -m\,V+i\,\Lambda\,U.  
\end{align}
In what follows, we clarify the apparent paradox of how energy is
conserved in Eqs.~\eqref{eq1}-\eqref{eq2} despite both components
experiencing gain or loss. It is straightforward to show that the
$\PTb$-symmetric Gross-Neveu model, represented by
Eqs.~\eqref{eq1}-\eqref{eq2}, can be transformed into 
\begin{align}
	\label{eq5}
	i \,(u_t-u_x)+(m-\Lambda)\,v -g|u\,v^\star+u^\star\,v|^{k} v &= 0, \\
	\label{eq6}
	i \,(v_t+v_x)+(m+\Lambda)\,u -g|u\,v^\star+u^\star\,v|^{k} u  &= 0,
\end{align}
and the new complex variables are defined as follows:
\begin{align}
	\label{eq3}
	u(x,t)&=\frac{V(-x,t)-i\,U(-x,t)}{\sqrt{2}}, \\
	\label{eq4}
	v(x,t)&=\frac{V(-x,t)+i\,U(-x,t)}{\sqrt{2}}. 
\end{align}
Writing the equations in this form offers several advantages:  
\begin{itemize}
	\item[(i)] It becomes clear that $\Lambda$ can be identified as
          the gain-loss parameter~\cite{alexeeva:2019}, since it appears 
	with opposite signs in Eqs.~\eqref{eq5}-\eqref{eq6}. Notice that
        this term has a different nature of dissipation introduced and 
	studied in Refs. \cite{quintero:2019a,quintero:2019b}.  
	\item[(ii)]  It can be easily shown that
          Eqs.~\eqref{eq5}-\eqref{eq6}, when expressed in light-cone
          coordinates $\chi = (x + t)/2$ and $\xi = (t - x)/2$, are
          invariant under the Lorentz transformation  
	\[
          \chi' = e^{-\beta} \chi,\quad \xi' = e^{\beta} \xi,\quad u' =
          e^{\beta/2} u,\quad v' = e^{-\beta/2} v,
	\]
	where the velocity of the moving frame is given by $V_s=\tanh(\beta)$.
\end{itemize}
The Lorentz transformations for the variables $U(x,t)$ and $V(x,t)$ are given by
\[
x' = \gamma(x - V_s t), \quad t' = \gamma(t - V_s x), \quad \text{and} \quad \Psi(x,t) = S\,\Psi'(x',t'),
\]
where the transformation matrix $S$ is
\[
S = \begin{pmatrix}
	\cosh(\beta/2) & i\sinh(\beta/2) \\
	-i\sinh(\beta/2) & \cosh(\beta/2)
\end{pmatrix}.
\]
\begin{itemize}
\item[(iii)] Eqs.~\eqref{eq5}-\eqref{eq6} are invariant under
  $\PTb$ symmetry, with parity and time-reversal transformations defined
  as
	\[
	\Pb: x \to -x,\ u \to v,\ v \to u, \qquad
	\Tb: t \to -t,\ u \to v^\star,\ v \to u^\star.
	\]
	These $\PTb$ symmetries imply the following relations:
	\begin{align}
		\label{eq:sym}
		u(x,t) &= u^\star(-x,-t), \qquad v(x,t) = v^\star(-x,-t).
	\end{align}	
In particular, the $\PTb$ symmetries ensure that if $(u(x, t), v(x,
t))^T$ is a solution of \eqref{eq5}-\eqref{eq6} for the associated
initial value $(u_0(x), v_0(x))^T$ then the transformed field
$(u^\star(-x, -t), v^\star(-x, -t))^T$ is also a solution of
\eqref{eq5}-\eqref{eq6} with the initial conditions $(u_0^\star(-x),
v_0^\star(-x))^T$. 	
\item[(iv)] If $(u(x, t), v(x, t))^T$ is a solution associated to the
  initial data $(u_0(x), v_0(x))^T$ for Eqs.~\eqref{eq5}-\eqref{eq6},
  then $(u^\star(x, t), -v^\star(x, t))^T$ is a solution associated to
  the initial data $(u_0^\star(x), -v_0^\star(x))^T$.  	
\item[(v)] Finally, the system of equations \eqref{eq5}-\eqref{eq6} is invariant 
under the following transformations:
\[
x \to -x,\ u \to -v,\ v \to -u,\ \Lambda \to -\Lambda,
\]
which implies that both components can bear either gain or loss, depending on the sign of $\Lambda$.
\end{itemize}

As shown, this representation makes it straightforward to obtain a stationary solution for
Eqs.~\eqref{eq5}-\eqref{eq6}. Additionally, a moving solitary wave can be found by applying
a Lorentz boost. 

The charge, canonical momentum, and energy densities are
defined as follows: 
\begin{align}
	\label{eq:dq}
\pazocal{{Q}}(x,t) &=|u(x,t)|^2+|v(x,t)|^2, \\
	\label{eq:dm}
\mathcal{P}_{c}(x,t) &= \frac{i}{2} (u_x\,u^\star-u_x^\star\,u+v_x\,v^\star-v_x^\star\,v),\\
	\label{eq:de}
	{\mathcal{E}}(x,t) &=\frac{i}{2}
                               (u_x\,u^\star-u_x^\star\,u-v_x\,v^\star+v_x^\star\,v)-m(u\,v^\star+u^\star\,v)\nonumber\\
  & +g\frac{|u\,v^\star+u^\star\,v|^{k}}{k+1}{(u\,v^\star+u^\star\,v)},	
\end{align}
respectively. From these definitions, the corresponding continuity equations  are given by
\begin{align}
	\label{eq:cq}
	\frac{\partial  {\pazocal{Q}}(x,t)}{\partial t} + \frac{\partial j(x,t)}{\partial x} &= 2 \,\Lambda\, i \,(u\,v^\star-u^\star\,v), \\
	\label{eq:cm0}
		\frac{\partial {\mathcal{P}_{c}}(x,t)}{\partial t} + \frac{\partial \Phi(x,t)}{\partial x} &= \Lambda\,(u\,v_x^\star+u^\star v_x-v\,u_x^\star-v^\star\,u_x), \\
	\label{eq:ce}
		\frac{\partial {\mathcal{E}}(x,t)}{\partial t} + \frac{\partial J(x,t)}{\partial x} &= 0,  
\end{align}
where
\begin{align}
	\label{eq:j}
j(x,t) &=|v|^2-|u|^2, \\
 \label{eq:Phi}
 \Phi(x,t) &=
                -m\,(u\,v^\star+u^\star\,v)+g\frac{|u\,v^\star+u^\star\,v|^{k}}{k+1} (u\,v^\star+u^\star\,v)\nonumber\\
               &+\frac{i}{2}(u\,u_t^\star-u^\star\,u_t+v\,v_t^\star-v^\star\,v_t),\\ 
  \label{eq:J}
  J(x,t) &=  \frac{i}{2}(u\,u_t^\star-u^\star\,u_t-v\,v_t^\star+v^\star\,v_t)+\Lambda\,(u\,v^\star+u^\star\,v).
\end{align}
By using the identity 
$$
u\,v_x^\star+u^\star\,v_x-v\,u_x^\star-v^\star\,u_x=-\frac{\partial}{\partial t} (u\,v^\star+u^\star\,v),
$$
Eq.\ \eqref{eq:cm0} can be written as follows
\begin{align}
	\label{eq:cm}
	\frac{\partial {\mathcal{P}}(x,t)}{\partial t} + \frac{\partial \Phi(x,t)}{\partial x} &= 0, 
\end{align}
where  ${\mathcal{P}}(x,t)=\mathcal{P}_{c}(x,t)+\Lambda\,
(u\,v^\star+u^\star\,v)$. While the canonical momentum
$P_{c}= \int \mathcal{P}_{c} \,dx$ is not conserved in general, clearly the momentum
\begin{align} \label{eq:mome}
	P &= \int {\cal{P}}\, dx
\end{align}
is conserved whenever $\Phi(+\infty,t)-\Phi(-\infty,t)=0$.

Assuming that \( j(+\infty,t) - j(-\infty,t) = 0 \), and \( J(+\infty,t)
- J(-\infty,t) = 0 \), it follows that the energy, 
\[
E = \int {\cal{E}}\, dx,
\]
is conserved. Moreover, the time derivative of the charge
\begin{align}
	\label{eq:timeQ}
	\frac{d Q}{d t} = 2 \,\Lambda\, i \,\int_{-\infty}^{+\infty} (u\,v^\star-u^\star\,v) \, dx, 
\end{align}
is apparently nonzero whenever $\Lambda \ne 0$. This agrees with the
time evolution of the charge obtained in Ref.~\cite{cuevas:2016} for the
case $k = 1$.  

Note that a nonlinear term, proportional to the coupling constant $g$,
appears only in the expressions for the energy density~\eqref{eq:de} and the momentum
flux~\eqref{eq:Phi}. This is a noteworthy observation, since the
relationship between these two quantities plays a key role in the
derivation of the exact solution. 

From the above definitions, we can derive two additional useful and simple identities:
\begin{align}
	\label{eq:c4}
	\mathcal{P}_{c}(x,t)&=-J(x,t), \\
	\label{eq:c5}
	{\cal{E}}(x,t)&=-\Phi(x,t) -m(u\,v^\star+u^\star v)-g\frac{k-1}{k+1} |u\,v^\star+u^\star\,v|^k (u\,v^\star+u^\star v).
\end{align} 
These identities will be used later to obtain the solution and to
compute the energy in a simpler way.  The last term in Eq.~\eqref{eq:c5}
appears due to nonlinear term in the NLD equation. This term vanishes when $k=1$.  


\section{Stationary and moving waves} \label{sec3}

Without loss of generality, the stationary solitary wave can be
represented by the ansatz~\cite{alexeeva:2019}
\begin{align}
	\label{eq:ansatza} 
	u(x,t) &= a(x)\,e^{i \theta(x)-i\omega t}, \\
	\label{eq:ansatzb} 
    v(x,t) &= -b(x)\,e^{i \varphi(x)-i\omega t},
\end{align}
where the functions $a(x) > 0$ and $b(x) > 0$ vanish as $x \to \pm \infty$. 	
We restrict our analysis to  $\omega>0$. Solutions for $\omega<0$ can be
directly obtained from this ansatz and the symmetry transformation (iv). 

The charge, canonical momentum and energy densities defined in
Eqs.~\eqref{eq:dq}-\eqref{eq:de} are now time-independent, and satisfy 
\begin{align}
	\label{eq:dq1}
	{\pazocal{Q}}(x) &=a^2(x)+b^2(x), \\
	\label{eq:dm1}
	\mathcal{P}_{c}(x) &=-a^2(x)\,\theta_{x}(x)-b^2(x)\,\varphi_{x}(x), \\ 
	\label{eq:de1}
  {\mathcal{E}}(x) &
  = -m\tau(x)+g \frac{|\tau(x)|^{k}}{k+1}\tau(x)+b^2(x)\,\varphi_{x}(x)-a^2(x)\,\theta_{x}(x), 
\end{align}
where 
\begin{align}
	\label{eq:sigma}
	\tau(x)&=u\,v^\star+u^\star v=-2\,a(x)\,b(x)\,\cos(\theta(x)-\varphi(x)).
\end{align}
Notice that $\tau(x)$ usually represents a scalar field in a 
$(1+1)$-dimensional field theory.

As a result, both the charge and the canonical momentum are  conserved
despite the presence of a source term in the corresponding continuity
equation. To proceed, we obtain the analytical expression of the
stationary solution. Moreover, since $\Pb$, $\mathcal{E}$, $J$, and
$\Phi$ are time-independent, not only are $\Pb$ and $\mathcal{E}$
conserved, but it also follows from Eqs.~\eqref{eq:ce} and \eqref{eq:cm}
that $J$ and $\Phi$ are constant as well. Due to the boundary
conditions, these two constants must be zero. As a consequence, from
Eq.\ \eqref{eq:c4}  the density of the canonical momentum
$\mathcal{P}_{c}$ is zero.

The continuity equation for charge \eqref{eq:cq}  leads to the following identity:
\begin{align}
	\label{eq:cj1}
	b\,b_x-a\,a_x &= 2\,a\,b \,\Lambda\,\sin(\theta(x)-\varphi(x)), 
\end{align}	
and the conditions $\Phi(x,t)=0$ and $J(x,t)=0$ yield the following algebraic relations 
\begin{align}
	\label{eq:ab1}
	m\tau(x) -g\frac{|\tau(x)|^{k}}{k+1}\tau(x)&=-\omega (a^2+b^2),\\
	\label{eq:ab2}
\Lambda \tau(x)&=-\omega (b^2-a^2),
\end{align}
respectively. 
Furthermore, the relations \eqref{eq:c4} and \eqref{eq:c5} reduce to
\begin{align}
	\label{eq:c4a}
	\mathcal{P}_{c}(x)&=-a^2(x)\,\theta_{x}(x)- b^2(x) \varphi_{x}(x)=0, \\
	\label{eq:c5a}
{\cal{E}}(x)&= -m\,\tau(x)-g \frac{k-1}{k+1}\,|\tau(x)|^{k}\tau(x).
\end{align} 
The first term in Eq.~(\ref{eq:c5a}) is a linear mass potential term. The parameter $m$ shifts the vacuum. The second term represents the self-interaction energy of the field and encodes fermion self-interactions. It is responsible for dynamical mass generation, and $g$ controls the strength of the interaction, as mentioned above.

\begin{figure}[h!]
	\begin{tabular}{c}
		\includegraphics[width=0.65\linewidth]{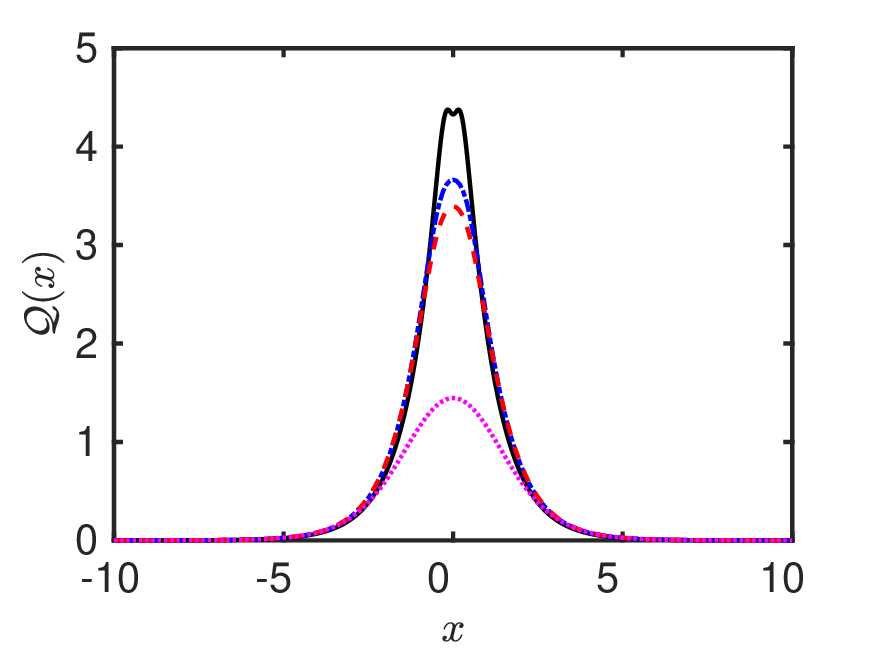} \\
		\includegraphics[width=0.65\linewidth]{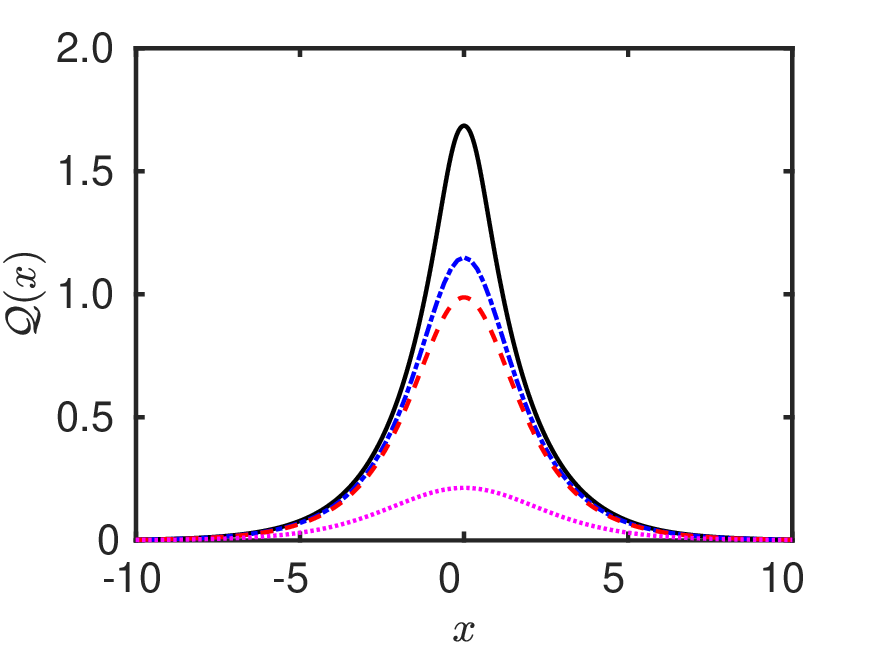} 
	\end{tabular}	
	\caption{Charge density	${\pazocal{Q}}(x)$ for $\omega=0.2$ (upper panel) and $\omega=0.5$ (lower panel). In both panels $k=1$ (magenta dotted line), $k=2.5$ (red dashed line), $k=3$ (blue dot-dashed line), 
		and $k=7$ (black solid line).  Other parameters: $m=1$, $g=1$, and $\Lambda=0.8$.}
	\label{fig1}
\end{figure}

From Eqs.~\eqref{eq:ab1} and \eqref{eq:ab2}, the functions $a(x)$ and
$b(x)$ can be expressed in terms of $\tau(x)$:
\begin{align}
	\label{eq:a2}
a^2(x)&=-\frac{\tau(x)}{2\omega} \left(-g \frac{|\tau(x)|^{k}}{k+1}+m-\Lambda\right),
\\
\label{eq:b2}
b^2(x)&=-\frac{\tau(x)}{2\omega} \left(-g\frac{|\tau(x)|^{k}}{k+1}+m+\Lambda\right).
\end{align}
The above two equations imply $\tau(x)<0$. Therefore, from the
definition \eqref{eq:sigma}, $\cos(\theta-\varphi)>0$. Taking into
account that $\tau(x) \to 0$ as $x \to \pm \infty$, these
expressions are well-defined provided that $|\Lambda| < m$.
From Eqs.~\eqref{eq:cj1} and \eqref{eq:ab2}, it follows that $\tau(x)$
satisfies the following differential equation: 
\begin{align}
	\label{eq:sigmax}
	\tau_{x}(x)&=-4\,a(x)\,b(x)\,\omega\,\sin(\theta(x)-\varphi(x)),
\end{align}
which is equivalent to 
\begin{align}
	\label{eq:sigmax2}
\tau_{x}(x)&= \pm 2 \tau(x) \left[\left(-g \frac{|\tau(x)|^{k}}{k+1}+m\right)^2-\rho^2\right]^{1/2},
\end{align}  
where $\rho^2 = \omega^2 + \Lambda^2$, and $ m^2>\rho^2$ (i.e., $0 <
\omega < \sqrt{m^2 - \Lambda^2}$). Due to the symmetry
$\tau(x)=\tau(-x)$, it is sufficient to consider the negative sign in
this equation, which holds for $x > 0$.   
This equation admits an exact solution given by:
\begin{align}
	\label{eq:solsigma}
	\tau(x)&=-\left(\frac{k+1}{2g}\right)^{1/k} 
	\left[
	\frac{2\,\kappa^2}{m+\rho \cosh(2k \kappa x)}
	\right]^{1/k},
\end{align}
where $\kappa = \sqrt{m^2 - \rho^2}$. For $k = 1$, this recovers the
function given in Eq.~(6.12) of Ref.~\cite{cuevas:2016}. By inserting
this solution into Eqs.~\eqref{eq:a2} and \eqref{eq:b2}, we obtain:
\begin{align}
	\label{eq:asol}
	a(x)&=\frac{1}{\sqrt{2\,\omega}}
	\left(\frac{(k+1)\,\kappa^2}{g(m+\rho\,\cosh[2k \kappa x])}\right)^{1/2k} 
	\sqrt{\frac{\rho^2-m\,\Lambda+(m-\Lambda)\,\rho\,\cosh(2k \kappa x)}{m+\rho\,\cosh(2k \kappa x)}},
	\\
	\label{eq:bsol}
	b(x)&= \frac{1}{\sqrt{2\,\omega}}
	\left(\frac{(k+1)\,\kappa^2}{g(m+\rho\,\cosh[2k \kappa x])}\right)^{1/2k} 
	\sqrt{\frac{\rho^2+m\,\Lambda+(m+\Lambda)\,\rho\,\cosh(2k \kappa x)}{m+\rho\,\cosh(2k \kappa x)}},
\end{align}
respectively. The soliton profile exhibits two humps whenever 
\begin{align}
	\label{eq:condition}
k & >\frac{\sqrt{\omega^2+\Lambda^2}}{m-\sqrt{\omega^2+\Lambda^2}}. 
\end{align}  
Otherwise, the soliton has a single-hump profile. 
This condition  implies that, for fixed parameters and a given frequency
$\omega$, there exists a threshold value of $k$ above which the soliton
profile develops two humps. This behavior is illustrated in
Fig.~\ref{fig1}, which shows the soliton 
profiles for different values of $k$ and frequencies $\omega$, with all
other parameters held constant. Specifically, fixing $m=1$,
$\Lambda=0.8$, for $\omega = 0.2$ (upper panel), solitary waves exhibit
a two-hump structure for values of $k$ such that $k \approx 4.7$,
whereas for $\omega = 0.5$ (lower panel), the critical value of $k$ for
the emergence of two humps is $k \approx 16.7$. The figure also shows
that the height of the single-hump profile increases with increasing
$k$. 
To complete our analysis, we now determine the phases $\theta(x)$ and $\varphi(x)$.  
Substituting the stationary ansatz into Eqs.~\eqref{eq:de} and \eqref{eq:c5}, and
 requiring that $\Phi(x,t)=0$, we find that
\begin{align}
	\label{eq:phases2}
	a^2(x)\,\theta_{x}(x)- b^2(x) \varphi_{x}(x)&=-\frac{g\,k}{k+1}|\tau(x)|^{k+1}\,.
\end{align} 
Together with Eq.~\eqref{eq:c4a}, this leads to the following two differential
equations satisfied by the phases:
\begin{align}
	\label{eq:thetax}
	\theta_{x}(x) &=-\frac{g\,k}{k+1}\frac{|\tau(x)|^{k+1}}{2\,a^2(x)}, \\
	\label{eq:varphix}
	 \varphi_{x}(x)&=\frac{g\,k}{k+1}\frac{|\tau(x)|^{k+1}}{2\,b^2(x)}. 
\end{align} 
The solutions to these equations (up to an additive constant $\Phi_0$) are given by:
\begin{align}
	\label{eq:thetasol}
	\theta(x) &=-\arctan\left(\frac{(m-\rho)(\rho+\Lambda)}{\omega \kappa} \tanh[k\,\kappa\,x]\right), \\
	\label{eq:varphisol}
	\varphi(x)&=\arctan\left(\frac{(m-\rho)(\rho-\Lambda)}{\omega \kappa} \tanh[k\,\kappa\,x]\right).
\end{align}
Therefore, the stationary solution of Eq.\ \eqref{eq5}-\eqref{eq6} are
represented by the functions $u(x,t)$ and $v(x,t)$ given by
Eqs.~\eqref{eq:ansatza}-\eqref{eq:ansatzb}, where the spatial parts are
defined by the functions \eqref{eq:asol}-\eqref{eq:bsol} and phases by
\eqref{eq:thetasol}-\eqref{eq:varphisol}. Notice that according to the
$\PTb$-symmetries of Eqs.~\eqref{eq5}-\eqref{eq6} if $(u(x, t), v(x,
t))^T$ is a stationary solution then $(u^\star(-x, -t), v^\star(-x,
-t))^T$ is also a stationary solution which might be different from the
original.

Once the exact stationary solution is obtained, its charge, momentum,
and energy can be calculated
\begin{align}
  \label{eq:charge}
  Q&= -\frac{g}{\omega (k+1)} I_k(k)+\frac{m}{\omega} I_0(k), \\
  \label{eq:momentum}
  P&=-\Lambda\,I_0(k), \\
  \label{eq:energy}
  E&= m\,I_0(k)+g \frac{k-1}{ k+1} I_k(k), 
\end{align}
where
\begin{align}\label{eq:ik}
	I_n(k) &\equiv \int_{-\infty}^{+\infty}|\tau(x)|^{n+1}\, dx 
	\\ \nonumber &
	=\frac{\sqrt{\pi}}{k \kappa}  \left[\frac{(k+1)(m-\rho)}{g}\right]^{\frac{n+1}{k}}\,
	\frac{\Gamma(\frac{n+1}{k})}{\Gamma(\frac{n+1}{k}+\frac{1}{2})}
                       \,_2F_1\left(\frac{n+1}{k},\frac{1}{2},\frac{n+1}{k}+\frac{1}{2},\alpha^2
                       \right) 
\end{align}
with $\alpha=\sqrt{(m-\rho)/(m+\rho)}$. For convenience, we explicitly
include the dependence of the integral  $I_n(k)$ on the parameter $k$.   
These expressions depend on the gain-loss coefficient $\Lambda$ through the parameters
$\rho=\sqrt{ \omega^2+\Lambda^2}$ and $\kappa=\sqrt{m^2-\rho^2}$.

The three magnitudes $Q$, $P$ and $E$ depend on $I_k(k)$ that encodes
the amount of nonlinearity. The charge $Q$ has two contributions: a
linear mass contribution that comes from the standard Dirac term and
a nonlinear correction that modifies the charge. Note that while the
canonical momentum is zero for stationary solution, the conserved
momentum $P \ne 0$ whenever $\Lambda \ne 0$.
There is an analogy with the minimal coupling in the standard
relativistic field theory (e.g. Quantum
Electrodynamics)~\cite{Griffiths:1987}, but now the momentum is
externally induced by $\Lambda$; no gauge fields and vector current
appear. The total energy $E$ splits into a linear term, proportional to
$I_0(k)$, a nonlinear interaction energy that is crucial in our analysis.

\begin{figure}[h!]
	\centering
	\includegraphics[width=0.65\linewidth]{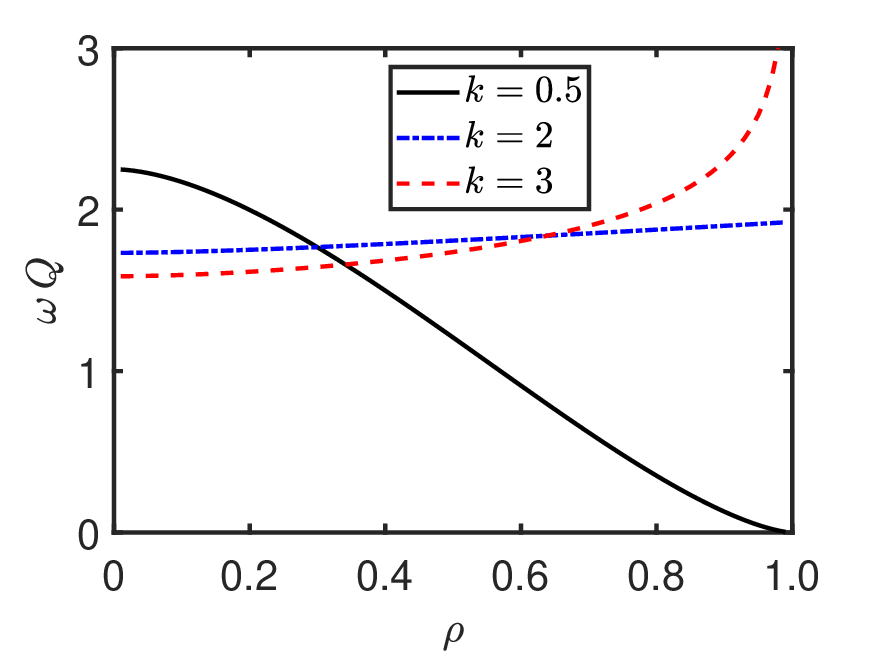} 	
	\caption{$\omega \,Q$ as  function of $\rho$ for different
          values of $k$. Other parameters are: $m=1$ and $g=1$.} 
	\label{fig2}
\end{figure}
To analyze the dependence of the charge on the nonlinearity parameter
$k$, Fig. \ref{fig2} illustrates $\omega \,Q$ as a function of
$\rho=\sqrt{\omega^2+\Lambda^2} \in (0,1)$ for $k=0.5$, $k=2$ and
$k=3$. The phase transition curve is reached at $\rho=1$. A significant
difference is observed at this boundary: while $\omega \, Q$ approaches
zero as $\rho \to 1$ for $ k<2$, it diverges to infinity for
$ k>2$ and achieves a constant value at $k=2$. This behavior can be
studied analytically using the expression  \eqref{eq:charge}. This
divergence contributes to a general trend where the maximum values of
$Q$ increase as $k$ gets larger. It is important to note, however, that
the overall behavior of $Q$ is not monotonic; their values can either
increase or decrease depending on the specific values of $\omega$ and
$\Lambda$. 

\subsection{Moving soliton solution}

From the stationary solution $u(x,t)$ and $v(x,t)$ and Lorentz
transformations, we can obtain the following moving soliton solution: 
\begin{align}
	\label{equ:ms1}
	u(x,t) &= e^{-\beta/2} a[\gamma(x-V_s t)] e^{i \theta[\gamma (x-V_s t)]} 
	e^{-i \omega \gamma (t-V_s x)}, \\ \label{eqv:ms1}
	v(x,t) &= -e^{\beta/2} b[\gamma(x-V_s t)] e^{i \varphi[\gamma (x-V_s t)]} 
	e^{-i \omega \gamma (t-V_s x)},
\end{align}
where $V_s=\tanh(\beta)$ is the soliton's velocity, and
$\gamma=\cosh(\beta)$ is the Lorentz factor. 

The charge, momentum, and energy of the moving soliton 
are given by 
\begin{align}
  \label{eq:charge1a}
  Q &= -\frac{g\,I_k(k)}{(1+k)\,\omega} +[m+\Lambda\,\tanh(\beta)] \frac{I_0(k)}{\omega},\\
  \label{eq:momentum1}
  P &= -\Lambda\,I_0(k) \,\cosh(\beta)- \left[m\,I_0(k)+g\, \frac{k-1}{
      k+1} I_k(k)\right] \sinh(\beta),\\
  \label{eq:energy1a}
  E &=\left[m\,I_0(k)+g\, \frac{k-1}{ k+1} I_k(k)\right] \,\cosh(\beta) + \Lambda\,I_0(k) \,\sinh(\beta), 
\end{align}
respectively. Interestingly, the expression \eqref{eq:charge1a}
demonstrates a direct dynamic coupling between the system's loss-gain
mechanism $\Lambda$ and its translational motion with a velocity
$\tanh(\beta)$.  

\begin{figure}[t!]
	\begin{tabular}{c}
		\includegraphics[width=0.7\linewidth]{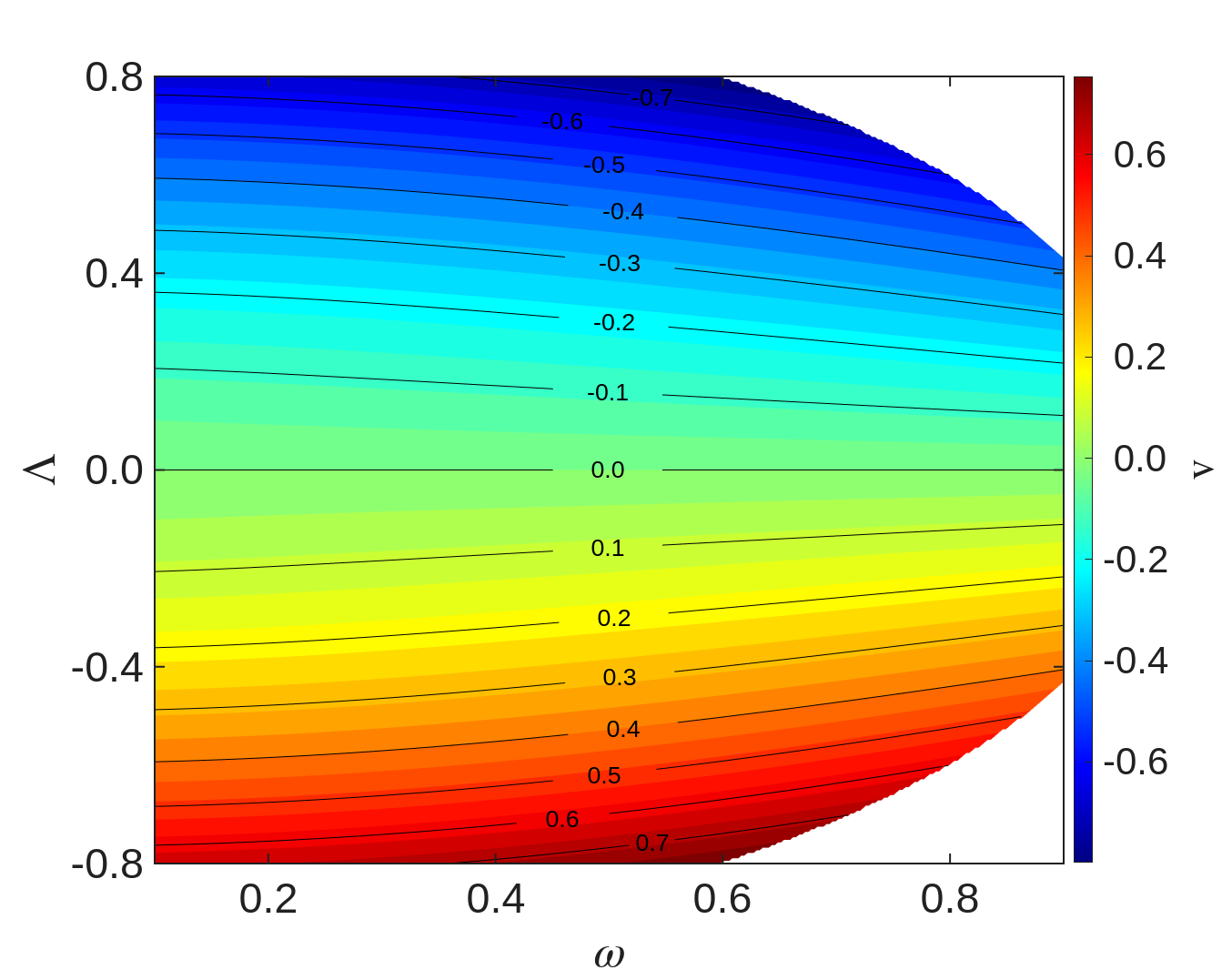} 
			\end{tabular}	
	\caption{Contour plots: soliton's velocity given by
          Eq.~\eqref{eq:vel} which guarantees zero momentum as
          functions of $\omega \in [0.1,0.9]$  and $\Lambda \in[-0.8,0.8]$.  
		Other parameters are: $m=1$,  $g=1$ and $k=3$.}
	\label{fig3}
\end{figure}
Furthermore, zero momentum can be achieved if the soliton's velocity satisfies
\begin{align}\label{eq:vel}
\tanh(\beta) \equiv \mathrm{v} &= \frac{-\Lambda\,I_0(k)}{m\,I_0(k)+g\,
                                 \frac{k-1}{ k+1} I_k(k)}
\end{align}
whenever  $|\mathrm{v}|<1$. For $k=1$, this condition is reduced to
$\mathrm{v}= -\Lambda$ and is independent of the soliton's frequency
$\omega$. In this case, the energy of the wave is
$E=I_0(1)/\cosh(\beta)$, whereas the energy of the moving Gross-Neveu
soliton scale as $\cosh(\beta)$. For other values of $k$, this specific
value of $v$ depends not only on $\Lambda$, but also on $\omega$. This
behaviour is shown in Fig.\ \ref{fig3} for $k=3$.
      
This result shows that the zero-momentum condition does not imply a
static  configuration. At this point, one can recall the case of a
relativistic particle in an electromagnetic field, where vanishing
momentum at nonzero velocity reflects the fact that the conserved
momentum differs from the mechanical momentum in the presence of an
external electromagnetic field. Thus, the total momentum can vanish even
when the particle is moving if the gauge field cancels out the
mechanical part~\cite{Griffiths:1987}.  

\section{Stability}\label{sec4}

In this section, we analyze the linear stability of the solitary wave
solutions~\eqref{eq:ansatza}-\eqref{eq:ansatzb}. We begin by linearizing
Eqs.~\eqref{eq5}-\eqref{eq6} around the stationary soliton 
defined in Eqs.~\eqref{eq:ansatza}-\eqref{eq:ansatzb} by introducing the
perturbed ansatz: 
\begin{equation}\label{stability_ansatz}
    \begin{aligned}
        \tilde{u} &= u(x, t)+ \varepsilon\zeta_1(x, t) = u(x, t)+ \varepsilon\hat{u}(x, t)e^{i\varphi-i\omega t}, \\
        \tilde{v} &=v(x, t)+ \varepsilon\zeta_2(x, t)= v(x, t)+ \varepsilon\hat{v}(x, t)e^{i\theta-i\omega t},
    \end{aligned}
\end{equation}
where $\varepsilon \ll 1$ is the perturbation amplitude. 
Substituting this ansatz into the governing equations
\eqref{eq5}-\eqref{eq6} and retaining terms up to the first order in
$\varepsilon$, we obtain the linearized problem: 
\begin{eqnarray}
\label{eq52}
i \,(\partial_t{\zeta_1}-\partial_x{\zeta_1})+(m-\Lambda)\,{\zeta_2} -g|\tau|^{k} {\zeta_2} +2gk |\tau|^{k-1} \Re\left(u^\star \zeta_2 + v^\star \zeta_1\right) v
&=& 0, \\
\label{eq62}
i \,(\partial_t{\zeta_2}+\partial_x{\zeta_2})+(m+\Lambda)\,{\zeta_1} -g|\tau|^{k} {\zeta_1} +2gk |\tau|^{k-1} \Re\left(u^\star \zeta_2 + v^\star \zeta_1\right) u
&=& 0,
\end{eqnarray}
or, equivalently,
\begin{equation}\label{linearized_system}
\partial_t y = \mathbf{L} y, 
\end{equation}
where $y = (\Re(\hat{u}), \Re(\hat{v}), \Im(\hat{u}), \Im(\hat{v}))^T$, and the linear operator $\mathbf{L}$ is given by:
\begin{equation}\label{matriz_L}
    \mathbf{L} =  J H, \qquad J = \begin{pmatrix}
        0 & I \\
        -I & 0 
    \end{pmatrix}, \qquad 
    H = \begin{pmatrix}
        A+B & -D_x \\
        D_x & -A
    \end{pmatrix}.
\end{equation}
The matrix differential operators $D_x$, $A$ and $B$ are explicitly defined as:
\begin{equation}
    \begin{aligned}
        D_x &= \gamma_0 (I \partial_x + \sin(\eta)M), & 
        A &=  \begin{pmatrix}
            \omega+\varphi_x & 0 \\
            0 & \omega-\theta_x 
        \end{pmatrix}+\cos(\eta)M,  \\
        B &=  2gk|\tau|^{k-1}\begin{pmatrix}
            b^2 & -ab \\
            -ab & a^2
        \end{pmatrix}, &
        M &= \begin{pmatrix}
            0 & (m-\Lambda)-g |\tau|^k \\
            (m+\Lambda)-g |\tau|^k & 0
        \end{pmatrix},
    \end{aligned}
\end{equation}
where the relative phase is defined as $\eta := \varphi-\theta$. Note
that $H$ is not self-adjoint in general, and consequently, the operator
$\mathbf{L}$ is not skew-adjoint. 

The spectral stability of the system described by
Eq. \eqref{linearized_system} relies entirely on the distribution of
eigenvalues $\lambda$ within $\sigma(\mathbf{L})$. A solitary wave is
classified as spectrally stable if all eigenvalues possess strictly
negative real parts ($\Re(\lambda) < 0$), and marginally stable if they
are non-positive ($\Re(\lambda) \leq 0$). Conversely, the emergence of
at least one eigenvalue with a strictly positive real part 
($\Re(\lambda) > 0$) renders the soliton spectrally unstable~\cite{berkolaiko:2012}.

We begin by examining the essential spectrum,
$\sigma_{\text{ess}}(\mathbf{L})$. Since the involved localized
functions decay exponentially, we can determine the essential spectrum
by evaluating the asymptotic limit of the operator $\mathbf{L}$ as $|x|
\to +\infty$ \cite{boussaid:2016}. By applying the Fourier 
transform to the limit operator, it is obtained that 
\begin{equation}
    \sigma_{\text{ess}}(\mathbf{L}) = i\mathbb{R}\setminus (-i\nu_1, i\nu_1),
\end{equation}
where $\nu_1 = \sqrt{m^2-\Lambda^2}-\omega$. Consequently, the essential
spectrum consists of two continuous phonon bands originating at the
branch points $\pm i\nu_1$  
and $\pm i\nu_2$, with $\nu_2 = \sqrt{m^2-\Lambda^2}+\omega$.

Now we turn our attention to the discrete spectrum,
$\sigma_{\text{d}}(\mathbf{L})$, which consists of isolated eigenvalues
with finite algebraic multiplicity. Here, we present certain
eigenfunctions that can be derived analytically through symmetry
considerations. 

Due to the continuous $U(1)$ gauge symmetry and the spatial
translational invariance of the governing Eq.~\eqref{eq5}, the operator
$\mathbf{L}$ possesses a nontrivial null space. Specifically, there
exist two localized zero-modes, $\phi_{\text{gau}}$ and
$\phi_{\text{tra}}$, belonging to $\ker(\mathbf{L})$, which take the explicit forms 
\begin{equation}
    \phi_{\text{gau}} = \begin{pmatrix}
        a \sin(\eta) \\
        b \sin(\eta) \\
        a \cos(\eta) \\
        -b \cos(\eta)
    \end{pmatrix}, \qquad     
    \phi_{\text{tra}} = \begin{pmatrix}
        a_x \cos(\eta)+ \theta_x a \sin(\eta) \\
        -b_x\cos(\eta)+ \varphi_x b \sin(\eta) \\
        -a_x\sin(\eta)+ \theta_x a \cos(\eta) \\
        -b_x \sin(\eta)-\varphi_x b \cos(\eta)
    \end{pmatrix}.
\end{equation}

By considering the exponential decay of the solutions at $\lambda= 0$, it can be shown that these two modes span the entire kernel, this is,
\begin{equation}
    \ker(\mathbf{L}) = \text{span}\{\phi_{\text{gau}} , \, \phi_{\text{tra}}\}.
\end{equation}
Consequently, the adjoint operator $\mathbf{L}^\dagger$ also admits a two-dimensional kernel spanned by the corresponding adjoint modes:
\begin{equation}
    \ker(\mathbf{L}^\dagger) = \text{span}\{\phi_{\text{gau}}^\dagger , \, \phi_{\text{tra}}^\dagger\}.
\end{equation}
While the underlying symmetries of the system guarantee the existence of
the geometric kernel for $\lambda = 0$, the algebraic multiplicity
strictly exceeds the geometric one due to the existence of a continuous
family of solutions, parameterized by the internal frequency $\omega$
and the boost velocity $V_s$. Differentiation with respect to these free
parameters and adapting them to the ansatz \eqref{stability_ansatz}
yields the generalized eigenfunctions $\phi_\omega$ and
$\phi_\mathrm{v}$ in the generalized null space, satisfying
$\mathbf{L}\phi_\omega = \phi_{\text{gau}}$ and
$\mathbf{L}\phi_\mathrm{v} = \phi_{\text{tra}}$. Explicitly, they read
\begin{equation}
    \phi_{\omega} = \begin{pmatrix}
        -a_\omega \cos(\eta)- \theta_\omega a \sin(\eta) \\
        b_\omega\cos(\eta)- \varphi_\omega b \sin(\eta) \\
        a_\omega\sin(\eta)- \theta_\omega a \cos(\eta) \\
        b_\omega \sin(\eta)+\varphi_\omega b \cos(\eta)
    \end{pmatrix}, \qquad 
    \phi_\mathrm{v} = \omega x\begin{pmatrix}
        a \sin(\eta) \\ b \sin(\eta) \\ a \cos(\eta) \\ -b \cos(\eta)
    \end{pmatrix}-\frac{1}{2}\begin{pmatrix}
        a \cos(\eta) \\ b \cos(\eta) \\ -a \sin(\eta) \\ b \sin(\eta)
    \end{pmatrix},
\end{equation}
which constitute the first level of a Jordan chain associated with the zero eigenvalue.

To extend the Jordan block structure further, one must find nontrivial
solutions $\phi$ and $\tilde{\phi}$ satisfying the inhomogeneous
equations: 
\begin{equation}\label{source}
    \mathbf{L} \phi = \phi_\omega, \qquad \mathbf{L} \tilde{\phi} = \phi_\mathrm{v}.
\end{equation}
According to the Fredholm alternative, these equations possess localized
solutions if and only if the independent terms of \eqref{source} are
orthogonal to the kernel of the adjoint operator. This yields the
following solvability conditions for each case: 
\begin{equation}
\begin{aligned}
    \text{VK}_1(\omega, \Lambda, k) &:= \sqrt{|\langle \phi_{\text{gau}}^\dagger | \phi_\omega \rangle|^2+|\langle \phi_{\text{tra}}^\dagger | \phi_\omega \rangle|^2} = 0, \\    
    \text{VK}_2(\omega, \Lambda, k) &:= \sqrt{|\langle \phi_{\text{gau}}^\dagger | \phi_\mathrm{v} \rangle|^2+|\langle \phi_{\text{tra}}^\dagger | \phi_\mathrm{v} \rangle|^2} = 0.
\end{aligned}
\end{equation}
Let us examine the limit where $\Lambda = 0$. In this special case, $H_{\Lambda = 0}$ becomes self-adjoint ($H = H^\dagger$), and thus $\mathbf{L}$ is skew-adjoint. 
The adjoint operator now takes the form $\mathbf{L}_{\Lambda = 0}^\dagger = - H J$. Therefore, its kernel is simply related to the kernel of $\mathbf{L}$ 
via the symplectic matrix $J$
\begin{equation}
    \ker(\mathbf{L}_{\Lambda = 0}^\dagger) = \text{span}\{J\phi_{\text{gau}}, J\phi_{\text{tra}}\}.
\end{equation}

Consequently, for $\Lambda = 0$, the solvability conditions reduce to
\begin{equation}
    \text{VK}_1(\omega, 0, k) := \left|\langle J\phi_{\text{gau}} | \phi_\omega \rangle\right| = \left|\frac{\partial Q}{\partial \omega}\right| = 0, \qquad  \text{VK}_2(\omega, 0, k) := |\langle J\phi_{\text{tra}} | \phi_\mathrm{v} \rangle | = E \neq 0,
\end{equation}
which recovers the well-known Vakhitov-Kolokolov (VK) criterion for
spectral stability, where $Q$ denotes the $U(1)$
charge~\cite{comech:2011}. 

Using the expression \eqref{eq:charge} of the charge $Q$ at $\Lambda =
0$, it can be shown (at least numerically) that no stability transition
point exists according to the VK criterion when $k \leq 2$. 
In contrast, for $k > 2$, there exists a critical frequency
$\omega_{\text{crit}}$ such that for $\omega_{\text{max}} =
\sqrt{m^2-\Lambda^2} > \omega > \omega_{\text{crit}}$ the soliton
becomes unstable, due to the emergence of a purely real eigenvalue in
the spectrum.

\begin{figure}[t]
	\centering
	\includegraphics[width=0.65\linewidth]{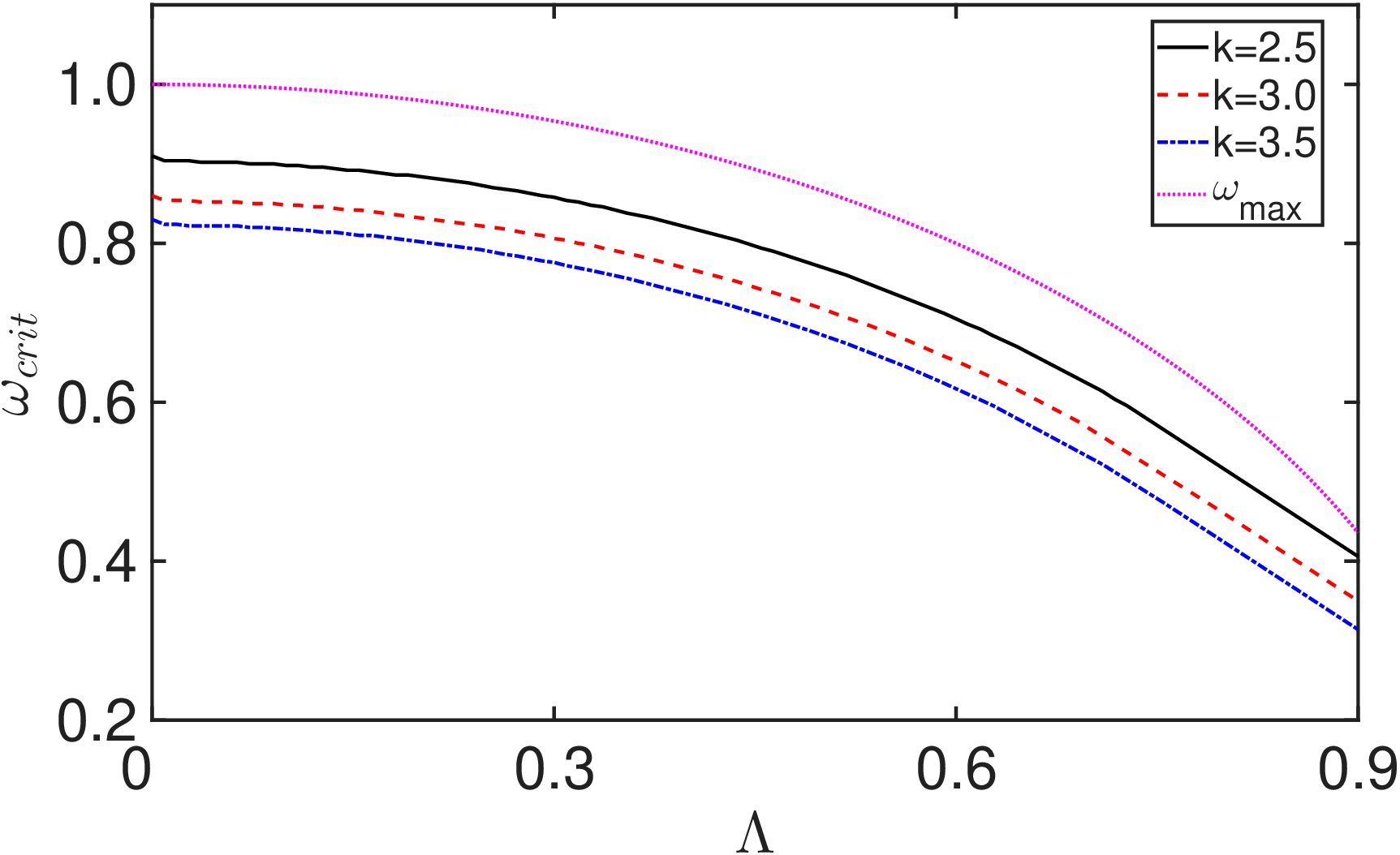}     
	\caption{Dependence of the critical frequency
          $\omega_{\text{crit}}$ on the gain-loss parameter $\Lambda$
          for several nonlinearity exponents ($k=2.5$, $3$, and
          $3.5$). The maximum frequency limit, $\omega_{\text{max}} =
          \sqrt{m^2-\Lambda^2}$, is included for reference. The fixed
          system parameters are $m=1$, $g=1$, $L=5$, and $N=901$.} 
	\label{fig4}
\end{figure}

\begin{figure}[htbp]
	\centering
	\includegraphics[width=0.65\linewidth]{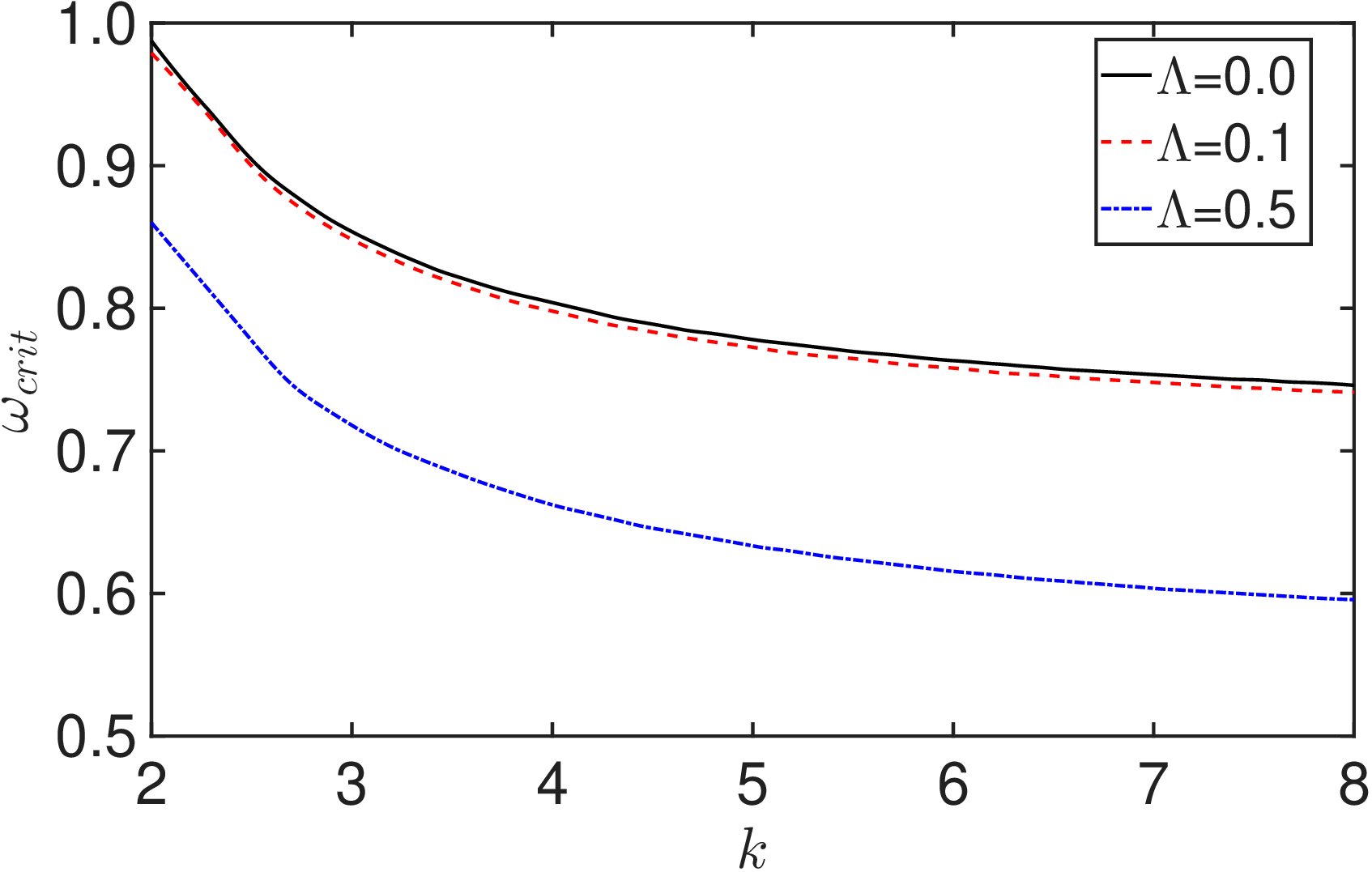}  
	\caption{Dependence of the critical frequency
          $\omega_{\text{crit}}$ on the nonlinearity exponent $k$ for
          selected values of the gain-loss parameter ($\Lambda=0$,
          $0.1$, and $0.5$). The fixed system parameters are $m=1$,
          $g=1$, $L=5$, and $N=901$.} 
	\label{fig5}
\end{figure}

In the general non-conservative case ($\Lambda \neq 0$), the symplectic
structure of $\mathbf{L}$ is broken, and thus no direct algebraic
relation exists between the operator and its adjoint form.  
To address this, we compute a numerical approximation of the critical
frequency $\omega_{\text{crit}}(\Lambda, k)$ through parametric sweeps
of the spectrum, as shown in Figs.~\ref{fig4} and~\ref{fig5}.

\begin{figure}[t]
    \centering
    \includegraphics[width=0.49\linewidth]{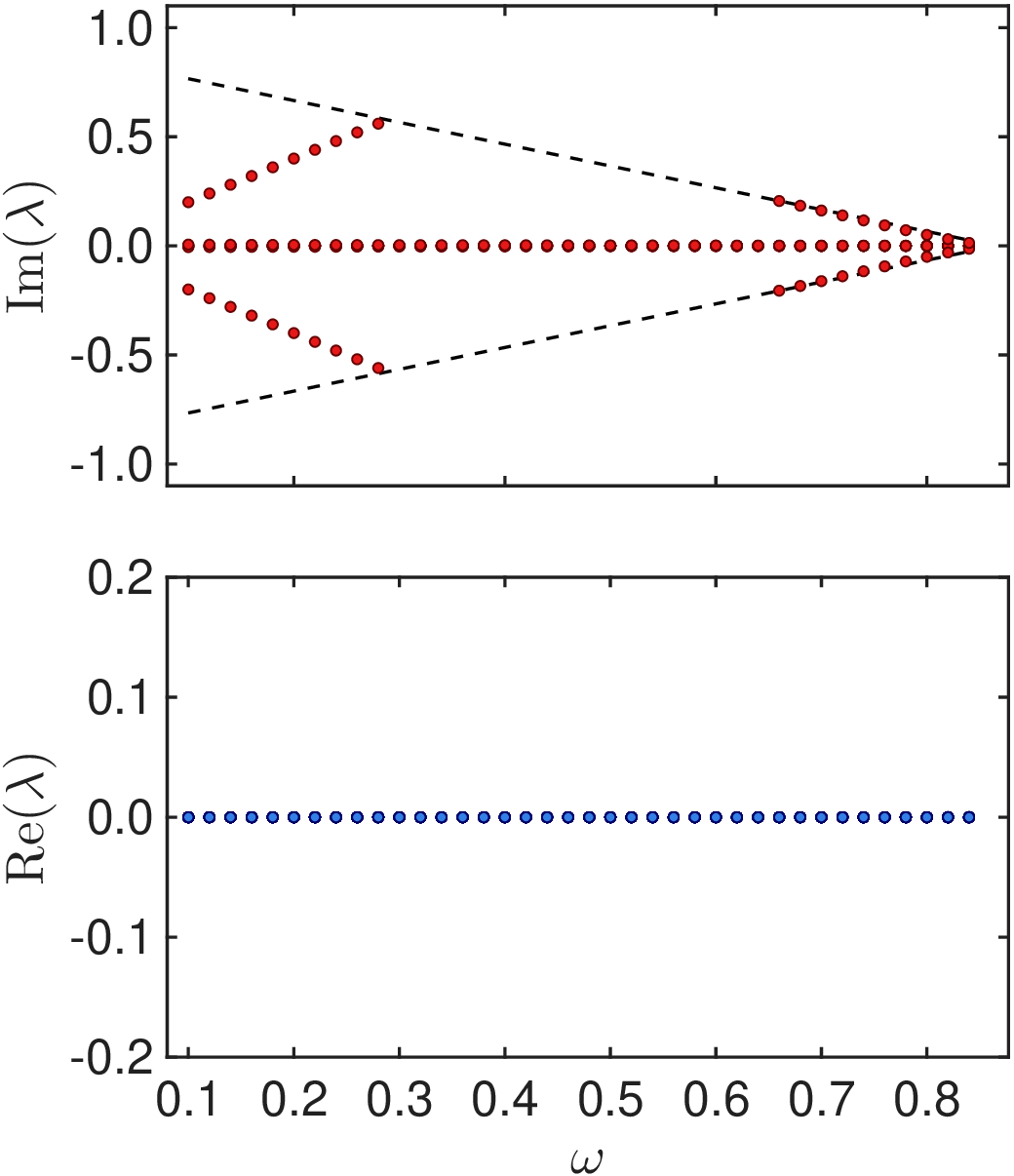}     
    \includegraphics[width=0.49\linewidth]{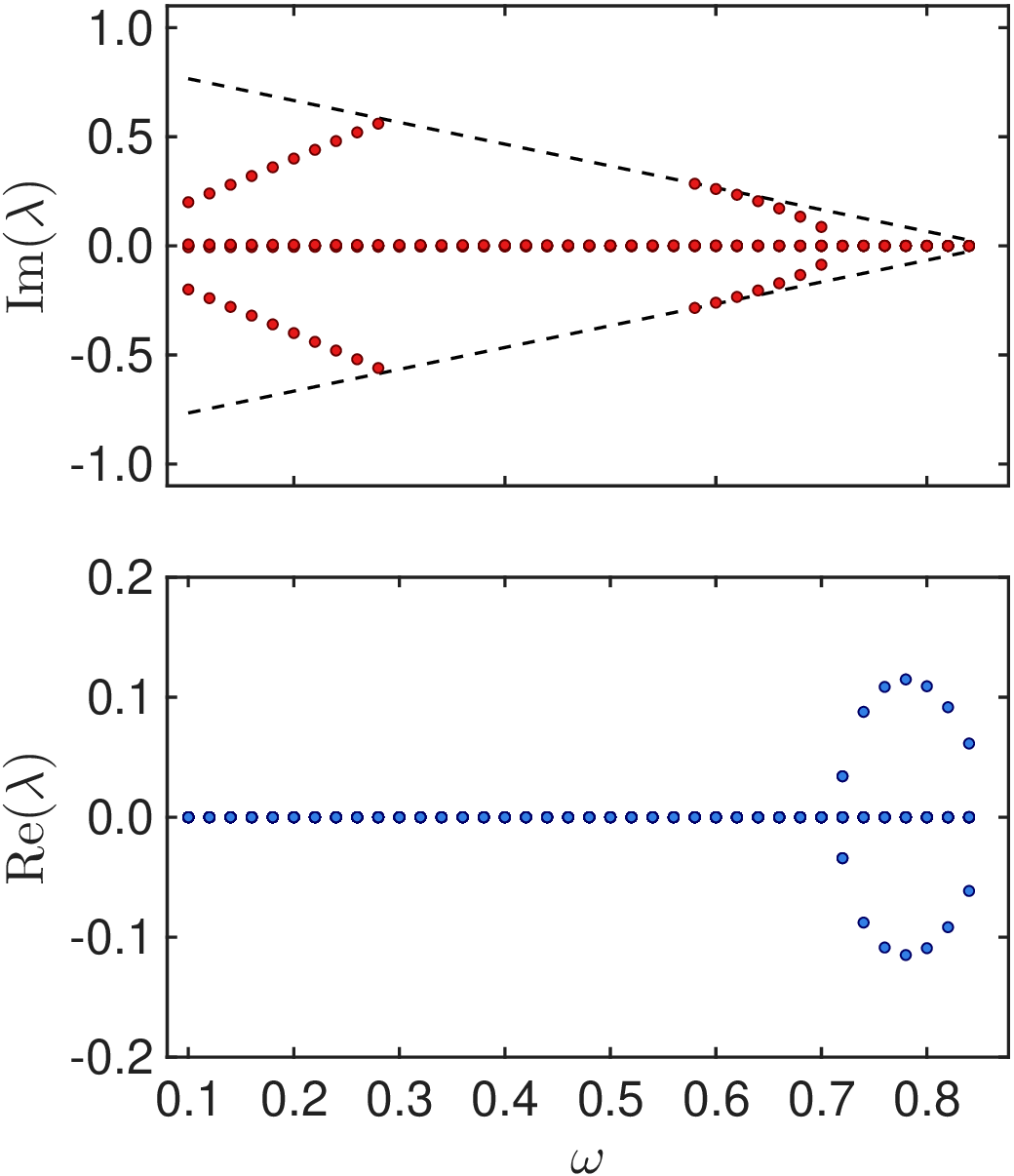} 
    \caption{Imaginary parts (top panels) and real parts (bottom panels) 
	of $\sigma_d(\mathbf{L})$ with respect to $\omega$ for a fixed $\Lambda=0.5$. 
	The left column corresponds to the nonlinearity exponent $k=2$
        ($\omega_{\text{crit}} = \sqrt{m^2-\Lambda^2} \approx 0.86$),  
while the right column corresponds to $k=3$ ($\omega_{\text{crit}} \approx 0.72$). 
Dashed lines represent the analytically predicted boundaries of the
essential spectrum. Computations were performed with $m=1$, $g=1$,
$L=5$, and $N=901$.} 
    \label{fig6}
\end{figure}

Our numerical analysis reveals that the critical frequency
$\omega_{\text{crit}}$ exists for $k > 2$. To illustrate this spectral
behavior, Fig.~\ref{fig6} displays the parametric evolution of both the
imaginary (top panels) and real (bottom panels) parts of the point
spectrum as a function of the frequency $\omega$, for a fixed gain-loss
parameter $\Lambda=0.5$ and representative nonlinearity exponents ($k=2$
and $3$). Note that for the particular case of $k=2$, the critical
frequency $\omega_{\text{crit}} = \omega_{\text{max}} =
\sqrt{m^2-\Lambda^2}$ which is the existence threshold of the soliton,
and thus no instability is observed in this case. 

To compute the stability spectrum, we utilize a Chebyshev spectral
collocation method combined with the algebraic mapping $x = L
\operatorname{atanh}(\xi)$\cite{chugunova:2006,chugunova:2007}.  
This transformation projects the standard Chebyshev nodes from the
interval $\xi \in (-1, 1)$ onto the entire real line $x \in (-\infty,
\infty)$, where the scaling factor $L$ optimizes resolution at the
soliton center and ensures the accurate capture of its asymptotic
decay. We achieve high spectral accuracy for the linearized  
operator $\mathbf{L}$ by employing $N=901$ collocation points and
setting $L=5$ for all numerical results presented. The corresponding
codes are available from the authors upon request.

As seen in Fig.~\ref{fig6}, a single pair of simple eigenvalues emerges
from the continuous spectrum and moves along the 
imaginary axis until they collide at zero for $\omega =
\omega_{\text{crit}}$. For $\omega > \omega_{\text{crit}}$, this pair of
eigenvalues becomes purely real and symmetric with respect to the
origin, thus rendering the soliton unstable in this regime. Crucially,
in the subcritical regime ($\omega < \omega_{\text{crit}}$) for $k \geq
2$, we explicitly note the absence of any complex eigenvalue quartets on
the spectral gap for the numerical computations which has been
performed; thus, the soliton remains spectrally marginally stable below
the critical threshold $\omega_{\text{crit}}$. Similarly, for $k < 2$,
no complex eigenvalue quartets are detected emerging from the spectral
gap at the frequencies studied, which implies that the solitary wave is
marginally stable.

As depicted in Fig.~\ref{fig4}, the critical frequency
$\omega_{\text{crit}}$ decreases as $\Lambda$ increases, suggesting that
the gain-loss mechanism exerts a strictly destabilizing effect on the
solitary waves. Furthermore, $\omega_{\text{crit}}$ also decreases as
$k$ increases (see Fig.~\ref{fig5}), indicating that solitons with
higher nonlinearity exponents are inherently more susceptible to
instability. 

Finally, another prominent feature of the spectrum of $\mathbf{L}$ is
the emergence of internal modes \cite{cuevas:2016} at $\lambda = \pm
2\omega i$. The existence of these discrete eigenvalues is deeply tied
to the underlying symmetry (iv) of Eqs.~\eqref{eq5}-\eqref{eq6}. To see
that, notice that $\zeta_1 = u^\star$ and $\zeta_2 = -v^\star$ is a
solution of the Eqs.~\eqref{eq52}-\eqref{eq62}. Therefore, we conclude
that the operator $\mathbf{L}$ in \eqref{linearized_system} satisfies 
\begin{equation}
\mathbf{L} \phi_{\pm} = \pm 2\omega i \,\phi_{\pm}, \qquad \phi_{\pm} = \begin{pmatrix}
   a \\ b \\ \mp ia  \\ \mp i b 
\end{pmatrix} e^{\mp i(\varphi+\theta)},
\end{equation}
i.e., it has $\lambda = \pm 2\omega i$ as eigenvalues, 
which can be clearly observed in the imaginary part of the spectrum
depicted in the top panels of Fig.~\ref{fig6}.

\section{Conclusions} \label{sec5}
 
We present the derivation of an exact solitary wave solution for the
$\PTb$-symmetric nonlinear Dirac equation involving a scalar-scalar
interaction with the generalized nonlinear term with an exponent $k$.
We established that the $\PTb$-transition point is defined solely by
the solution's existence condition $\omega^2+\Lambda^2<m^2$, and is
independent of the nonlinearity exponent $k$. By using the continuity
equations of charge, energy and momentum, we obtain the exact
analytical stationary solution, and we demonstrate that energy is
conserved, despite the presence of a gain-loss term characterized by the
parameter $\Lambda$. A key non-trivial result is that the stationary
solution, obtained analytically, possesses a nonzero momentum when the
gain-loss parameter $\Lambda$ is nonzero. This unusual momentum in the
rest frame highlights the peculiar nature of the system. 

We also derived the moving soliton solution by applying a Lorentz boost
to the stationary solution. We explicitly calculated the charge
($Q$), energy ($E$), and momentum ($P$) of the moving solution as
functions of the soliton's frequency, velocity, and the gain-loss
parameter. We showed that the gain-loss parameter $\Lambda$ acts as a
crucial control parameter, allowing the soliton's velocity to be
precisely selected such that its total momentum vanishes. This
capability offers a  dynamic control mechanism in $\PTb$-symmetric
nonlinear systems. Furthermore, we confirmed that the energy and
momentum of the moving soliton satisfy the fundamental relativistic mass
relation, $E^2-P^2=M_0^2$, where the rest mass $M_{0}$ depends on the
nonlinear exponent $k$, on the soliton's frequency, and on the gain-loss
parameter.  

Beyond finding exact solutions, we analyzed their linear spectral
stability. In the conservative limit ($\Lambda=0$), we analytically
recovered the Vakhitov-Kolokolov (VK) criterion. For the general
$\PTb$-symmetric case ($\Lambda \neq 0$), our numerical results revealed
a clear dependence on the nonlinearity $k$. Solitary waves with $k < 2$
are marginally stable. However, for $k \geq 2$, a critical frequency
$\omega_{\text{crit}}$ emerges: the soliton is marginally stable for
$\omega < \omega_{\text{crit}}$ and becomes unstable for $\omega >
\omega_{\text{crit}}$ due to the appearance of a purely real
eigenvalue. Furthermore, both the gain-loss mechanism ($\Lambda$) and
higher nonlinearities ($k$) destabilize the system by lowering this
critical frequency. In summary, our work provides a comprehensive
analytical framework for understanding the solitary wave solutions of
the $\PTb$-symmetric Gross-Neveu equation. 
 
\section*{Acknowledgments}
F.C.N. was supported by 
FPU24/00951 (Ministerio de Ciencia, Innovación y Universidades, Spain). 
S.P. is partially supported by Spanish MINECO/FEDER Grants
PGC2022-126078NB-C21 
and PID2024-160228NB-I00, funded by MCIN/AEI/10.13039/\ 501100011033 and
“ERDF A way of making Europe”, and Grant E21-23R funded by Arag\'on
Government and the European Union, and the NextGenerationEU Recovery
and Resilience Program on {\it{Astrof\'isica y F\'isica de Altas
    Energ\'ias}} CEFCA-CAPA-ITAINNOVA.
R.A.N. was partially supported by PID2024-155593NB-C21 (FEDER(EU)/ Ministerio de Ciencia, Innovación y Universidades-Agencia Estatal de Investigación)
 DOI: 10.13039/501100011033,
Spain). F.C.N., R.A.N. and N.R.Q. were supported by FQM-415 (Plan
Propio, Universidad de Sevilla, Spain) and IMUS-Maria de Maeztu grant CEX2024-001517-M - Apoyo a Unidades de Excelencia María de Maeztu for supporting this research, funded by MICIU/AEI/ DOI: 10.13039/501100011033.

\section*{ORCID iDs}
Fernando Carreño-Navas  https://orcid.org/0009-0004-4819-1472 \\
 Siannah Peñaranda-Rivas  https://orcid.org/0000-0002-9408-4406  \\
 Renato Alvarez-Nodarse https://orcid.org/0000-0002-9038-4698\\
 Niurka R Quintero https://orcid.org/0000-0003-3503-3040
 
%
 
 
\end{document}